\documentclass[12pt,preprint]{aastex}

\usepackage{amsmath,graphicx,sidecap}

\begin{document}
\bibliographystyle{natbib}
\thispagestyle{empty}
\newcommand{\h}{$^{\rm h}$}
\newcommand{\m}{$^{\rm m}$}
\newcommand{\s}{$^{\rm s}$}
\newcommand{\rsun}{$R_{\odot}$}
\newcommand{\msun}{$M_{\odot}$}
\newcommand{\lsun}{$L_{\odot}$}
\newcommand{\teff}{$T_{\rm eff}$}
\newcommand{\adross}{$\theta_{\rm Ross}$}
\newcommand{\adud}{$\theta_{\rm UD}$}
\newcommand{\kbar}{$\bar{k}_{2obs}$}

\begin{center}
{\bf  \large Fundamental Stellar Astrophysics Revealed at Very High Angular Resolution}
\end{center}

\vskip 4 in

\begin{tabular}{l}
{\bf Contact: Jason Aufdenberg}\\
{\bf (386) 226-7123}\\
{\bf Embry-Riddle Aeronautical University, Physical Sciences Department}\\
{\bf aufded93@erau.edu} \\
\\
{\bf Co-authors:}\\
{\bf Stephen Ridgway (National Optical Astronomy Observatory)}\\
{\bf Russel White  (Georgia State University)}\\
\end{tabular}

\newpage
\pagestyle{myheadings}
\markright{Fundamental Stellar Astrophysics Revealed at Very High Angular Resolution}
\setcounter{page}{1}

\section*{Introduction}

A detailed understanding of stellar structure and evolution is vital
to all areas of astrophysics.  In exoplanet studies the age and mass
of a planet are known only as well as the age and mass of the hosting
star, mass transfer in intermediate mass binary systems lead to type
Ia Supernova that provide the strictest constraints on the rate of the
universe's acceleration, and massive stars with low metallicity and
rapid rotation are a favored progenitor for the most luminous events
in the universe, long duration gamma ray bursts.  Given this universal
role, it is unfortunate that our understanding of stellar astrophysics
is severely limited by poorly determined basic stellar properties -
effective temperatures are in most cases still assigned by blunt
spectral type classifications and luminosities are calculated based on
poorly known distances.  Moreover, second order effects such as rapid
rotation and metallicity are ignored in general.  Unless more
sophisticated techniques are developed to properly determine
fundamental stellar properties, advances in stellar astrophysics will
stagnate and inhibit progress in all areas of astrophysics.
Fortunately, over the next decade there are a number of observational
initiatives that have the potential to transform stellar astrophysics
to a high-precision science.  Ultra-precise space-based photometry
from CoRoT (2007+) and Kepler (2009+) will provide stellar seismology
for the structure and mass determination of single stars. GAIA (2011+)
will yield precise distances to nearly a billion stars, providing
accurate luminosities.  However, \textit{the unprecedented data from
  these upcoming missions will only translate to useful calibrations
  of stellar models if they are performed in concert with high angular
  resolution measurements provided by ground-based optical and
  infrared interferometers.}

Very high angular resolution (reaching $<$ 1 milliarcsecond)
observational astronomy at optical and infrared wavelengths is still
in its infancy.  Only in the last decade have multi-element
interferometers become fully operational.  In this white paper we
highlight some of the incredible achievements made with optical and
infrared interferometry over the last decade, and use these to
emphasize the potential for these facilities, with continued support,
to transform stellar astrophysics into a high-precision science.

\section*{Accurate Stellar Masses}

Arguably the most fundamental parameter for a star is its mass, as this 
sets the timescale for evolution and determines its ultimate fate.  
Unfortunately mass is a difficult property to measure directly, and is
typically only available for stars in binary systems with observable orbital 
motion.  Mass estimates for the generic single stars must rely on 
mass-temperature or mass-luminosity calibrations, or worse, the predictions 
of untested stellar evolutionary models.  The high resolution 
capabilities of optical interferometry have the potential to greatly 
increase the number and types of stars for which we have dynamical mass
estimates, and greatly improve overall mass estimates.

The recent discoveries of young stars in nearby moving groups such as
the $\beta$ Pictoris Association and the TW Hydrae Association
\cite[e.g.][]{zs2004} provide many
new opportunities to determine dynamical masses of young binaries
\cite[e.g.][]{boden2005,boden2007,sspb2008}.  Mass estimates at this early age are
especially important because of the poorly understood input physics
(e.g. convection, opacities) of pre-main sequence stars.  With precise
distance estimates to be provided by GAIA, relative orbits will
provide dynamical masses of stars in rarer evolutionary states, such
as those transitioning to the giant phases \cite[e.g.][]{bth05}, traversing the Hertzsprung Gap
\cite[]{btl06}, and
high-mass main sequence and Wolf Rayet stars \cite[]{north07,kraus07}.
 Only just recently have the prospects of high
contrast imaging via non-redundant aperture masking on large-diameter
telescopes been realized.  This work can provide dynamical masses for
substellar objects \cite[e.g.]{ireland08}, a
mass range where evolutionary models are very poorly constrained due
to the age/tempertaure/mass degeneracies.  As very high resolution
interferometric techniques continue to mature, becoming more adaptable
and sensitive, they will become an essential tool for determining
stellar masses for stars spanning the entire H-R diagram.

\subsection*{Asteroseismology and Interferometry}
While binary stars will continue to be the dominant device to
determine stellar masses, density measurements provided by powerful
asteroseismology measurements combined with radius measurements by
optical interferometry offer a method to estimate masses for single
stars.  The photometric oscillations observed by asteroseismology
reveal interior stellar properties through their dependence on the
density (sound speed) distribution within the star.  Since such
distributions are a function of both mass and age, both of these
fundamental stellar parameters are probed.  Reliable masses and ages
from asteroseismology require tight constraint of global stellar
parameters, most importantly the stellar radius. Stellar radii from
interferometry accurate to 3\% when coupled with asteroseismology
yield single star masses accurate to 4\% \cite[]{cunha07}.  In the
case of the nearby subgiant $\beta$ Hydri, this has been done to a
precision of 2.8\% \cite[]{north07B}. Other examples of connections
between asteroseismology and interferometry include a check on the
mass of a $\delta$ Scuti star in the Hyades \cite[]{armstrong06} and a
radius and age for asteroseismology target $\tau$ Ceti (G8 V)
\cite[]{difolco04}.  The synergy between ground-based interferometry
and both ground-based programs (HARPS, CORALIE, ELODIE, UVES, UCLES,
SIAMOIS, SONG) and spaced-based (MOST, CoRoT, WIRE, Kepler, PLATO)
high-precision photometric missions will play a vital role in
determining fundamental stellar masses.

\begin{figure}
\includegraphics*[scale=1.0]{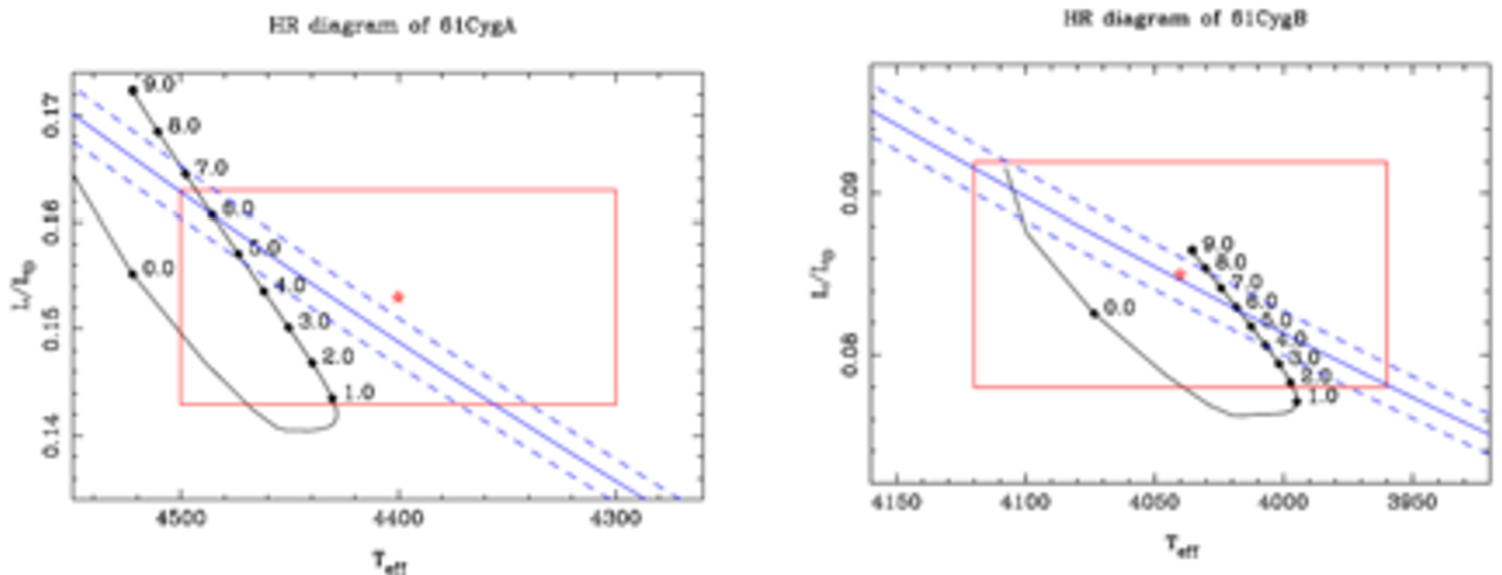}
\caption{Evolutionary tracks in the H-R diagram
for 61 Cyg A ({\bf left}) and B ({\bf right}). The labels indicate the age in Gyr
relative to the ZAMS. The rectangular box represents the classical
$L-T_{\rm eff}$ error box, and the diagonal lines represent the radius
and its uncertainty determined by a combination of interferometry and
the Hipparcos parallax \cite[]{kervella08}}
\label{61cyg}
\end{figure}

\section*{Precise Radii and Temperatures}
With available mass estimates, astronomers can begin the next step in
high precision stellar astrophysics - testing the temperature and size
predictions of stellar evolutionary models as a function of age.
Historically the only stars that astronomers could measure accurate
radii and (relative) temperatures of were in eclipsing binary systems
\cite[]{andersen91}.  Long-baseline optical/infrared
interferometry has changed this dramatically by providing sufficient
resolution to measure the angular size of many nearby stars.  
Angular measurements with errors well under 1\% are now commonplace. 
The limb-darkening corrections, normally modeled, are now subject to
direct interferometric verification.  

In order to convert a stellar angular diameter into a physical
diameter, the distance must be known.  Accurate binary orbits can
provide this in some cases, but  a more general solution is
increasingly possible, with Hipparcos, GAIA and SIM
providing/promising 1\% distances to 10, 500 or 2500 pc. 
Interferometric + Hipparcos measurements of single M-dwarfs show that
they are 10-15\% larger than currently predicted by models
\cite[]{berger08}, with a suggestion that the discrepancy increases
with elevated metallicity.  Examples of ultra-precise radii and
temperatures (see Figure~\ref{61cyg}) have been measured for coeval
binary stars \cite[]{kervella08}, metal poor population II stars \cite[e.g.][]{boy08}, 
giant stars in the Hyades \cite[]{boy09}, and the enigmatic $\lambda$
Boo itself \cite[]{ciardi07}.

\begin{SCfigure}
\includegraphics[width=0.6\textwidth]{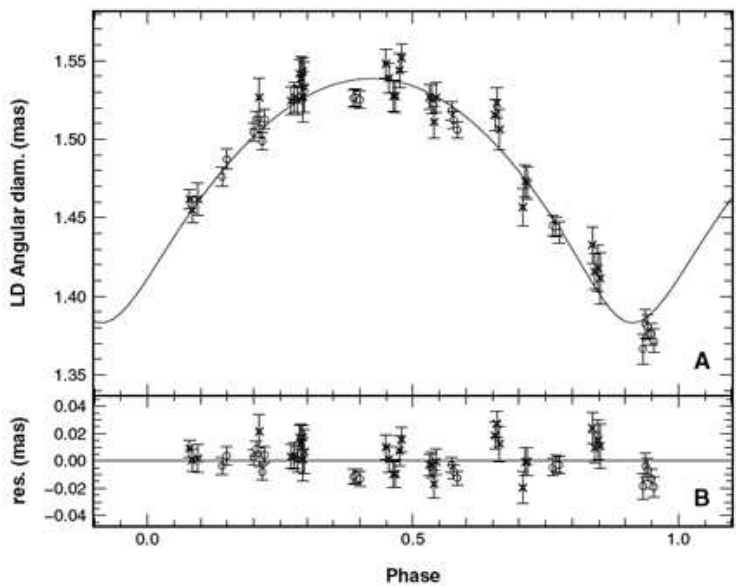}
\caption{Angular diameter measurements showing the 5.36 day pulsation
  of $\delta$ Cephei with a model fit (A) and residuals (B)
\cite[]{merand05}. The solid curve is
  predicted from radial velocity data, where only the amplitude of the
  curve is adjusted to determine the projection factor.}
\label{deltacep}
\end{SCfigure}

Regarding stars with variable radii, a great achievement during the
last decade has been the interferometric measurement of the pulsating
diameters of seven classical cepheids: $\zeta$ Gem \cite[]{lane2000},
$\eta$ Aql, W Sgr, $\beta$ Dor, l Car \cite[]{kervella04}, $\delta$
Cep \cite[]{merand05}, and Y Oph \cite[]{merand07}.  The precision of
the $\delta$ Cep (see Figure~\ref{deltacep}) and Y Oph measurements
was such that the projection factor, a value normally predicted by
models and needed to correct radial velocity measurements, could be
observationally constrained.  GAIA distances will extend detailed
study to a broad range of Cepheids.

The \teff\ of a star is one of the small number of fundamental
parameters of a stellar model. In order to confront models with real
stars, the corresponding description of the star must be known. It is
defined in terms of the luminosity and the radius by

\[ T_{\rm eff} = \biggl[\frac{L}{4\pi R^2\sigma}\biggr]^{1/4} 
= \biggl[\frac{4 f_{\rm bol}}{\theta_{\rm LD}\sigma}\biggr]^{1/4} \]
where $f_{\rm bol}$ is the bolometric flux, $\theta_{\rm LD}$ is the
angular diameter corrected for limb darkening, and $\sigma$ is the
Stefan-Boltzmann constant.

\begin{center}
\begin{figure}
\includegraphics*[scale=1.0]{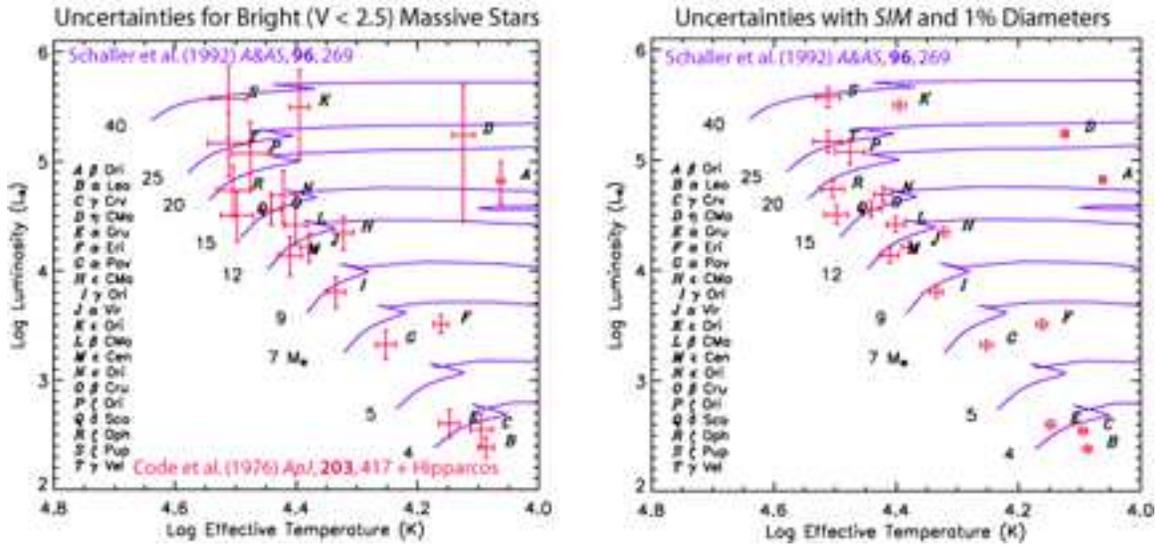}
\caption{{\bf Left:} The original Hanbury Brown intensity
interferometer diameters together with the Hipparcos parallaxes,
compared to massive star evolutionary tracks. {\bf Right:} Expected
error bars with SIM parallaxes and optical amplitude interferometry
angular diameters, compared to evolutionary tracks.}
\label{hrplot}
\end{figure}
\end{center}

For a classical star with a well-defined surface, this lets us
determine the \teff\ by measuring the angular diameter and the
observed flux, and modulo an understanding of any extinction, gives a
precise answer.  The determination of \teff\ is typically limited by
the photometry (or, with good photometry, by the absolute calibration
of the photometry). In fact, with accumulating interferometric stellar
measurements, it is now possible to turn the question around and, from
stellar photometry alone, predict the angular diameters and \teff\ of
common spectral types to ~1-2\% \cite[]{kervella04B}.  In the case of
stars with a poorly defined surface (e.g. due to extended atmosphere,
accretion disk, mass loss shell), an imaging capability allows a more
detailed confrontation of observed and modeled brightness
distributions \cite[e.g.][]{perrin04,wittkowski08} In the case of very
hot stars, \teff\ is particularly difficult to determine owing to the
inability to observe the turnover of the Planck function in the far
ultraviolet. With a measured angular diameter, a single ophotometric
measure can give the surface brightness, and thanks to the simplicity
of the spectrum, a reliable \teff, and with distance a reliable
luminosity, see Figure~\ref{hrplot}.

\section*{Limb Darkening}
While precise masses and radii obtained from interferometric
observations test and constrain stellar structure and evolution
models, very high resolution observations of stellar photospheres test
stellar atmosphere models, models that are vital for the construction
of synthetic stellar spectral energy distributions and high-resolution
synthetic spectra.  Measurement of the center-to-limb intensity
variation of a stellar photosphere probes the temperature structure of
that atmosphere.  This was first done for the Sun just over 100 years
ago and helped to constrain models for the transport of energy in
the solar atmosphere.  Such studies are now possible for other stars.

Until recently, with the exception of Sirius (A1 V)
\cite{hb74}(Hanbury Brown, R. et al 1974, MNRAS, 167,475),
interferometric limb-darkening measurements have been limited to a
small number of stars cooler than the sun: $\alpha$ Cas (K0 III) and
$\alpha$ Ari (K2 III) \cite[]{hajian98}; Arcturus (K1.5 III)
\cite[]{quirrenbach96}; Betelgeuse (M1Iab) \cite[]{perrin04B};
$\gamma$ Sagittae (M0 III), V416 Lac (M4 III), and BY Boo (M4.5 III)
\cite[]{wittkowski01}; $\psi$ Phoenicis (M4 III) \cite[]{psi_phe04}.

\begin{SCfigure}
\includegraphics[width=0.70\textwidth]{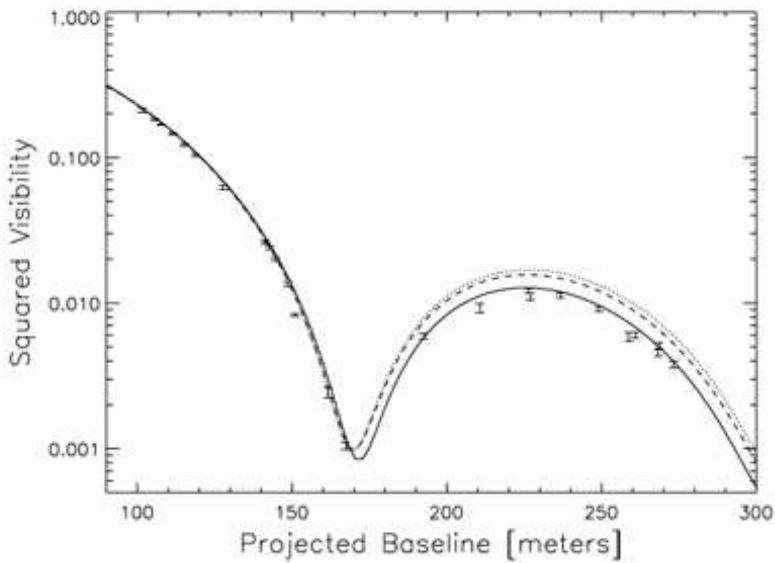}
\caption{A visibility curve for Vega (A0 V) showing the first and
  second lobes.  The first lobe yields the angular size, while the
  second lobe contains limb-darkening information.  In the case of
  Vega, a pole-on rapid rotator, the data show both limb and gravity
  darkening (solid line) \cite[]{vega06}. Such data are vital for
  probing the temperature structures of stellar atmospheres.}
\label{vega}
\end{SCfigure}

In the last five years as long baselines have come on-line, hotter
stars have been measured: Altair (A7 V) \cite[]{ohishi04}, Vega (A0 V)
\cite[]{vega06, vega06_peterson}, $\alpha$ Cyg (A2 Ia) and $\beta$ Ori
(B8 Ia) \cite[]{aufdenberg08}.  The interferometric confirmation that
Vega is a pole-on, rapidly rotating star (see Figure~\ref{vega}) has
recently driven astronomers to search for a replacement star, or set
of stars, for defining photometric systems \cite[]{rieke08}.

\subsection*{Limb Darkening and Convection}
High precision multi-wavelength angular diameter measurements reveal
the wavelength dependence of limb-darkening.  Such measurements of
Procyon \cite[]{procyon05} and $\alpha$ Cen B \cite[]{bigot06} provide
tests of 3-D convective transport models.  The interferometric data
indicate a temperature gradient shallower than provided by the
standard 1-D mixing-length convection.  Further limb-darkening
measurements, in particular at high spectral resolution (up to
R$\sim$30,000), will allow for more stringent tests and constraints on
the state-of-the-art multi-dimensional stellar atmosphere codes, those
responsible for the recent substantial revision in the Sun's oxygen
abundance \cite[]{asplund04}.  {\em Global convective instabilities
  are present either in the core or envelope of most stars.
  Nevertheless, the appropriate modelling of convection remains one of
  the most difficult tasks in the context of stellar astrophysics.} -
\cite{cunha07}.  High-resolution studies of stellar surfaces will play
an important role in testing improved models for convection as they
develop.
\section*{Stellar Rotation}

\begin{figure}[h!]
\includegraphics*[scale=1.0]{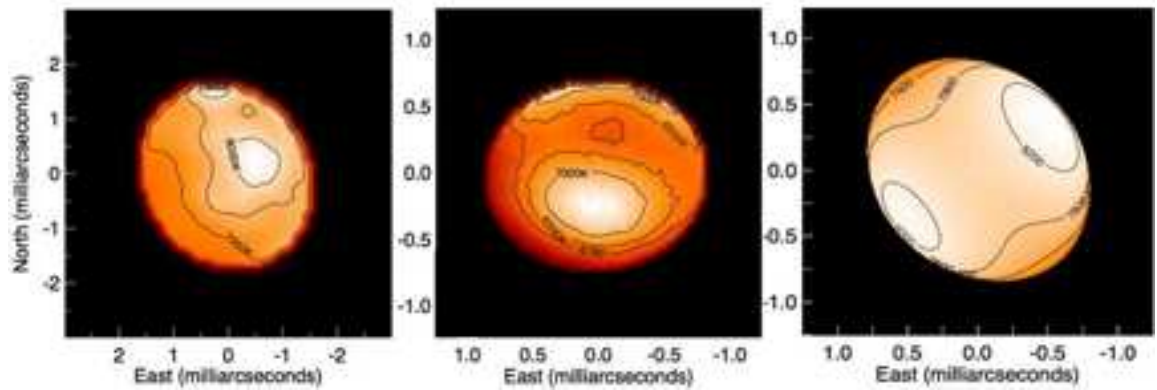}
\caption{Model-independent images of rapid rotator stars $\alpha$ Aql
  ({\bf left:} \cite{monnier07} and $\alpha$
  Cep ({\bf middle:} Zhao et al. 2009, in preparation) showing the
  shape effects of centripetal forces, and brightness distributions
  determined by polar brightening and equatorial darkening, polar axis
  projected angle, and limb darkening. {\bf Right:} A Roche-von
  Zeipel model for $\alpha$ Oph based on interferometric data (Zhao et
  al. 2009, in preparation.)}
\label{mirc}
\end{figure}

Rotating stars offer a powerful tool for insight into stellar
interiors. The mass distribution, opacities, and processes such as
differential rotation and convection are normally lost in the
spherical uniformity of normal stars.  In rapid rotators, these
factors contribute to the stellar shape and temperature distribution,
which may follow more or less closely the idealizations of Roche and
von Zeipel, and these factors will be implicated differently for stars
of differing mass. A range of rotation rates for similar stars
effectively constitutes a series of experiments.  Interferometric
imaging shows the distorted stellar shape.  A first measurement of
rotational deformation in $\alpha$ Eri \cite[]{dds03} has already
stimulated 14 publications of follow-up or interpretation.  More
recent imaging shows also the distribution of brightness temperature
across the disk, and the limb darkening (eg. Zhao et al. 2009, in
preparation, see Figure~\ref{mirc}).  For rapid rotators, of which Be
stars are the classic example, interferometry can map the ejected
material, which may be found in disks \cite[]{tycner06} and/or polar
winds \cite[]{kervella09}.  Imaging may also serve an important role
in understanding the interaction of rotation and pulsation, as in for
example $\delta$ Sct stars \cite[]{peterson06} and other non-radial
oscillators \cite[]{jankov01}.

Recent observations have spurred theoreticians to replace the Roche
formalism (which uses a potential where the mass is a point source)
with a more physically realistic model including a self-consistent
gravitational potential and differential rotation \cite[]{jackson05,
  mac07}.  These models predict that rotation significantly reduces
the luminosity of young stars, by reducing the central core
temperature, resulting, for example, in a 1.2 \msun\ star with a
luminosity of 1.0 \lsun.  Such models make matters worse for resolving
the faint young Sun paradox \cite[e.g.][]{minton07}, where the
standard solar model predicts the young Sun too cool to support life
on the early Earth 4.0 Gyr ago, in contradiction to the geologic and
fossil records.  Testing the latest models of rapidly rotating stars
with interferometric imaging will have an impact across stellar
astrophysics.

\section*{High Angular Resolution Imaging in the Next Decade}

Interferometric techniques already cover a large parameter space.  In
the near-IR, angular resolutions reach $<$1 milliarcsec, with
single-measurement precision as fine as 20 microarcsec.  Wavelength
coverage extends over most of the atmospheric windows from 0.48 to
12.5 $\mu$m.  Spectral resolutions are available up to ~30,000 in R
and I, ~12,000 in J-H-K, adequate for characterizing most molecular
bands and some individual spectral lines.  In the N band, spectral
resolutions up to ~200 are well matched to grain and molecular opacity
structure.  Instrumentation and facility developments already underway
will augment these capabilities somewhat, improve limiting
sensitivity, and so forth.  But the most dramatic development will be
in the quality of imaging achieved.  Most interferometric measurements
are still made with only one telescope pair.  Beam combination with 3,
4 or 6 telescopes is just in its infancy, but will rapidly come to be
the preferred observing mode for the kinds of science described above,
owing to its far more rapid accumulation of image information.
Interferometry has important complementary potential for space, as
well, for missions such as SIM, SI (Stellar Imager), and SPIRIT (Space
Infrared Interferometric Telescope).  The coming decade will be a
period of rich development for high resolution stellar science.

\bibliography{white_paper}

\begin{thebibliography}{}

\bibitem[{Andersen}(1991){Andersen}]{andersen91}
{Andersen}, J. (1991).
\newblock {Accurate masses and radii of normal stars}.
\newblock {\em \aapr\/}, {\bf 3}, 91--126.

\bibitem[{Armstrong} {\em et~al.}(2006){Armstrong}, {Mozurkewich}, {Hajian},
  {Johnston}, {Thessin}, {Peterson}, {Hummel}, and {Gilbreath}]{armstrong06}
{Armstrong}, J.~T., {Mozurkewich}, D., {Hajian}, A.~R., {Johnston}, K.~J.,
  {Thessin}, R.~N., {Peterson}, D.~M., {Hummel}, C.~A., and {Gilbreath}, G.~C.
  (2006).
\newblock {The Hyades Binary ${\theta}^{2}$ Tauri: Confronting Evolutionary
  Models with Optical Interferometry}.
\newblock {\em \aj\/}, {\bf 131}, 2643--2651.

\bibitem[{Asplund} {\em et~al.}(2004){Asplund}, {Grevesse}, {Sauval}, {Allende
  Prieto}, and {Kiselman}]{asplund04}
{Asplund}, M., {Grevesse}, N., {Sauval}, A.~J., {Allende Prieto}, C., and
  {Kiselman}, D. (2004).
\newblock {Line formation in solar granulation. IV. [O I], O I and OH lines and
  the photospheric O abundance}.
\newblock {\em \aap\/}, {\bf 417}, 751--768.

\bibitem[{Aufdenberg} {\em et~al.}(2005){Aufdenberg}, {Ludwig}, and
  {Kervella}]{procyon05}
{Aufdenberg}, J.~P., {Ludwig}, H.-G., and {Kervella}, P. (2005).
\newblock {On the Limb Darkening, Spectral Energy Distribution, and Temperature
  Structure of Procyon}.
\newblock {\em \apj\/}, {\bf 633}, 424.

\bibitem[{Aufdenberg} {\em et~al.}(2006){Aufdenberg}, {M{\'e}rand}, {Foresto},
  {Absil}, {Di Folco}, {Kervella}, {Ridgway}, {Berger}, {Brummelaar},
  {McAlister}, {Sturmann}, {Sturmann}, and {Turner}]{vega06}
{Aufdenberg}, J.~P., {M{\'e}rand}, A., {Foresto}, V.~C.~d., {Absil}, O., {Di
  Folco}, E., {Kervella}, P., {Ridgway}, S.~T., {Berger}, D.~H., {Brummelaar},
  T.~A.~t., {McAlister}, H.~A., {Sturmann}, J., {Sturmann}, L., and {Turner},
  N.~H. (2006).
\newblock {First Results from the CHARA Array. VII. Long-Baseline
  Interferometric Measurements of Vega Consistent with a Pole-On, Rapidly
  Rotating Star}.
\newblock {\em \apj\/}, {\bf 645}, 664--675.

\bibitem[{Aufdenberg} {\em et~al.}(2008){Aufdenberg}, {Ludwig}, {Kervella},
  {M{\'e}rand}, {Ridgway}, {Coud{\'e} Du Foresto}, {Ten Brummelaar}, {Berger},
  {Sturmann}, and {Turner}]{aufdenberg08}
{Aufdenberg}, J.~P., {Ludwig}, H.-G., {Kervella}, P., {M{\'e}rand}, A.,
  {Ridgway}, S.~T., {Coud{\'e} Du Foresto}, V., {Ten Brummelaar}, T.~A.,
  {Berger}, D.~H., {Sturmann}, J., and {Turner}, N.~H. (2008).
\newblock {Limb Darkening: Getting Warmer}.
\newblock In A.~{Richichi}, F.~{Delplancke}, F.~{Paresce}, and A.~{Chelli},
  editors, {\em The Power of Optical/IR Interferometry: Recent Scientific
  Results and 2nd Generation\/}, pages 71--82.

\bibitem[{Berger} {\em et~al.}(2008){Berger}, {ten Brummelaar}, {Gies},
  {Henry}, {McAlister}, {Merand}, {Sturmann}, {Sturmann}, {Turner},
  {Aufdenberg}, and {Ridgway}]{berger08}
{Berger}, D.~H., {ten Brummelaar}, T.~A., {Gies}, D.~R., {Henry}, T.~J.,
  {McAlister}, H.~A., {Merand}, A., {Sturmann}, J., {Sturmann}, L., {Turner},
  N.~H., {Aufdenberg}, J.~P., and {Ridgway}, S.~T. (2008).
\newblock {The Radius-Luminosity Relation from Near-Infrared Interferometry:
  New M Dwarf Sizes from the CHARA Array}.
\newblock In G.~{van Belle}, editor, {\em 14th Cambridge Workshop on Cool
  Stars, Stellar Systems, and the Sun\/}, volume 384 of {\em Astronomical
  Society of the Pacific Conference Series\/}, pages 226--+.

\bibitem[{Bigot} {\em et~al.}(2006){Bigot}, {Kervella}, {Th{\'e}venin}, and
  {S{\'e}gransan}]{bigot06}
{Bigot}, L., {Kervella}, P., {Th{\'e}venin}, F., and {S{\'e}gransan}, D.
  (2006).
\newblock {The limb darkening of {$\alpha$} Centauri B. Matching 3D
  hydrodynamical models with interferometric measurements}.
\newblock {\em \aap\/}, {\bf 446}, 635--641.

\bibitem[{Boden} {\em et~al.}(2005a){Boden}, {Sargent}, {Akeson}, {Carpenter},
  {Torres}, {Latham}, {Soderblom}, {Nelan}, {Franz}, and
  {Wasserman}]{boden2005}
{Boden}, A.~F., {Sargent}, A.~I., {Akeson}, R.~L., {Carpenter}, J.~M.,
  {Torres}, G., {Latham}, D.~W., {Soderblom}, D.~R., {Nelan}, E., {Franz},
  O.~G., and {Wasserman}, L.~H. (2005a).
\newblock {Dynamical Masses for Low-Mass Pre-Main-Sequence Stars: A Preliminary
  Physical Orbit for HD 98800 B}.
\newblock {\em \apj\/}, {\bf 635}, 442--451.

\bibitem[{Boden} {\em et~al.}(2005b){Boden}, {Torres}, and {Hummel}]{bth05}
{Boden}, A.~F., {Torres}, G., and {Hummel}, C.~A. (2005b).
\newblock {Testing Stellar Models with an Improved Physical Orbit for 12
  Bootis}.
\newblock {\em \apj\/}, {\bf 627}, 464--476.

\bibitem[{Boden} {\em et~al.}(2006){Boden}, {Torres}, and {Latham}]{btl06}
{Boden}, A.~F., {Torres}, G., and {Latham}, D.~W. (2006).
\newblock {A Physical Orbit for the High Proper Motion Binary HD 9939}.
\newblock {\em \apj\/}, {\bf 644}, 1193--1201.

\bibitem[{Boden} {\em et~al.}(2007){Boden}, {Torres}, {Sargent}, {Akeson},
  {Carpenter}, {Boboltz}, {Massi}, {Ghez}, {Latham}, {Johnston}, {Menten}, and
  {Ros}]{boden2007}
{Boden}, A.~F., {Torres}, G., {Sargent}, A.~I., {Akeson}, R.~L., {Carpenter},
  J.~M., {Boboltz}, D.~A., {Massi}, M., {Ghez}, A.~M., {Latham}, D.~W.,
  {Johnston}, K.~J., {Menten}, K.~M., and {Ros}, E. (2007).
\newblock {Dynamical Masses for Pre-Main-Sequence Stars: A Preliminary Physical
  Orbit for V773 Tau A}.
\newblock {\em \apj\/}, {\bf 670}, 1214--1224.

\bibitem[{Boyajian} {\em et~al.}(2008){Boyajian}, {McAlister}, {Baines},
  {Gies}, {Henry}, {Jao}, {O'Brien}, {Raghavan}, {Touhami}, {ten Brummelaar},
  {Farrington}, {Goldfinger}, {Sturmann}, {Sturmann}, {Turner}, and
  {Ridgway}]{boy08}
{Boyajian}, T.~S., {McAlister}, H.~A., {Baines}, E.~K., {Gies}, D.~R., {Henry},
  T., {Jao}, W.-C., {O'Brien}, D., {Raghavan}, D., {Touhami}, Y., {ten
  Brummelaar}, T.~A., {Farrington}, C., {Goldfinger}, P.~J., {Sturmann}, L.,
  {Sturmann}, J., {Turner}, N.~H., and {Ridgway}, S. (2008).
\newblock {Angular Diameters of the G Subdwarf {$\mu$} Cassiopeiae A and the K
  Dwarfs {$\sigma$} Draconis and HR 511 from Interferometric Measurements with
  the CHARA Array}.
\newblock {\em \apj\/}, {\bf 683}, 424--432.

\bibitem[{Boyajian} {\em et~al.}(2009){Boyajian}, {McAlister}, {Cantrell},
  {Gies}, {Brummelaar}, {Farrington}, {Goldfinger}, {Sturmann}, {Sturmann},
  {Turner}, and {Ridgway}]{boy09}
{Boyajian}, T.~S., {McAlister}, H.~A., {Cantrell}, J.~R., {Gies}, D.~R.,
  {Brummelaar}, T.~A.~t., {Farrington}, C., {Goldfinger}, P.~J., {Sturmann},
  L., {Sturmann}, J., {Turner}, N.~H., and {Ridgway}, S. (2009).
\newblock {Angular Diameters of the Hyades Giants Measured with the CHARA
  Array}.
\newblock {\em \apj\/}, {\bf 691}, 1243--1247.

\bibitem[{Ciardi} {\em et~al.}(2007){Ciardi}, {van Belle}, {Boden}, {ten
  Brummelaar}, {McAlister}, {Bagnuolo}, {Goldfinger}, {Sturmann}, {Sturmann},
  {Turner}, {Berger}, {Thompson}, and {Ridgway}]{ciardi07}
{Ciardi}, D.~R., {van Belle}, G.~T., {Boden}, A.~F., {ten Brummelaar}, T.,
  {McAlister}, H.~A., {Bagnuolo}, Jr., W.~G., {Goldfinger}, P.~J., {Sturmann},
  J., {Sturmann}, L., {Turner}, N., {Berger}, D.~H., {Thompson}, R.~R., and
  {Ridgway}, S.~T. (2007).
\newblock {The Angular Diameter of {$\lambda$} Bo{\"o}tis}.
\newblock {\em \apj\/}, {\bf 659}, 1623--1628.

\bibitem[{Cunha} {\em et~al.}(2007){Cunha}, {Aerts}, {Christensen-Dalsgaard},
  {Baglin}, {Bigot}, {Brown}, {Catala}, {Creevey}, {Domiciano de Souza},
  {Eggenberger}, {Garcia}, {Grundahl}, {Kervella}, {Kurtz}, {Mathias},
  {Miglio}, {Monteiro}, {Perrin}, {Pijpers}, {Pourbaix}, {Quirrenbach},
  {Rousselet-Perraut}, {Teixeira}, {Th{\'e}venin}, and {Thompson}]{cunha07}
{Cunha}, M.~S., {Aerts}, C., {Christensen-Dalsgaard}, J., {Baglin}, A.,
  {Bigot}, L., {Brown}, T.~M., {Catala}, C., {Creevey}, O.~L., {Domiciano de
  Souza}, A., {Eggenberger}, P., {Garcia}, P.~J.~V., {Grundahl}, F.,
  {Kervella}, P., {Kurtz}, D.~W., {Mathias}, P., {Miglio}, A., {Monteiro},
  M.~J.~P.~F.~G., {Perrin}, G., {Pijpers}, F.~P., {Pourbaix}, D.,
  {Quirrenbach}, A., {Rousselet-Perraut}, K., {Teixeira}, T.~C.,
  {Th{\'e}venin}, F., and {Thompson}, M.~J. (2007).
\newblock {Asteroseismology and interferometry}.
\newblock {\em \aapr\/}, {\bf 14}, 217--360.

\bibitem[{Di Folco} {\em et~al.}(2004){Di Folco}, {Th{\'e}venin}, {Kervella},
  {Domiciano de Souza}, {Coud{\'e} du Foresto}, {S{\'e}gransan}, and
  {Morel}]{difolco04}
{Di Folco}, E., {Th{\'e}venin}, F., {Kervella}, P., {Domiciano de Souza}, A.,
  {Coud{\'e} du Foresto}, V., {S{\'e}gransan}, D., and {Morel}, P. (2004).
\newblock {VLTI near-IR interferometric observations of Vega-like stars. Radius
  and age of {$\alpha$} PsA, {$\beta$} Leo, {$\beta$} Pic, {$\epsilon$} Eri and
  {$\tau$} Cet}.
\newblock {\em \aap\/}, {\bf 426}, 601--617.

\bibitem[{Domiciano de Souza} {\em et~al.}(2003){Domiciano de Souza},
  {Kervella}, {Jankov}, {Abe}, {Vakili}, {di Folco}, and {Paresce}]{dds03}
{Domiciano de Souza}, A., {Kervella}, P., {Jankov}, S., {Abe}, L., {Vakili},
  F., {di Folco}, E., and {Paresce}, F. (2003).
\newblock {The spinning-top Be star Achernar from VLTI-VINCI}.
\newblock {\em \aap\/}, {\bf 407}, L47--L50.

\bibitem[{Hajian} {\em et~al.}(1998){Hajian}, {Armstrong}, {Hummel}, {Benson},
  {Mozurkewich}, {Pauls}, {Hutter}, {Elias}, {Johnston}, {Rickard}, and
  {White}]{hajian98}
{Hajian}, A.~R., {Armstrong}, J.~T., {Hummel}, C.~A., {Benson}, J.~A.,
  {Mozurkewich}, D., {Pauls}, T.~A., {Hutter}, D.~J., {Elias}, II, N.~M.,
  {Johnston}, K.~J., {Rickard}, L.~J., and {White}, N.~M. (1998).
\newblock {Direct Confirmation of Stellar Limb Darkening with the Navy
  Prototype Optical Interferometer}.
\newblock {\em \apj\/}, {\bf 496}, 484--+.

\bibitem[{Hanbury Brown} {\em et~al.}(1974){Hanbury Brown}, {Davis}, {Lake},
  and {Thompson}]{hb74}
{Hanbury Brown}, R., {Davis}, J., {Lake}, R.~J.~W., and {Thompson}, R.~J.
  (1974).
\newblock {The effects of limb darkening on measurements of angular size with
  an intensity interferometer}.
\newblock {\em \mnras\/}, {\bf 167}, 475--484.

\bibitem[{Ireland} {\em et~al.}(2008){Ireland}, {Kraus}, {Martinache}, {Lloyd},
  and {Tuthill}]{ireland08}
{Ireland}, M.~J., {Kraus}, A., {Martinache}, F., {Lloyd}, J.~P., and {Tuthill},
  P.~G. (2008).
\newblock {Dynamical Mass of GJ 802B: A Brown Dwarf in a Triple System}.
\newblock {\em \apj\/}, {\bf 678}, 463--471.

\bibitem[{Jackson} {\em et~al.}(2005){Jackson}, {MacGregor}, and
  {Skumanich}]{jackson05}
{Jackson}, S., {MacGregor}, K.~B., and {Skumanich}, A. (2005).
\newblock {On the Use of the Self-consistent-Field Method in the Construction
  of Models for Rapidly Rotating Main-Sequence Stars}.
\newblock {\em \apjs\/}, {\bf 156}, 245--264.

\bibitem[{Jankov} {\em et~al.}(2001){Jankov}, {Vakili}, {Domiciano de Souza},
  and {Janot-Pacheco}]{jankov01}
{Jankov}, S., {Vakili}, F., {Domiciano de Souza}, Jr., A., and {Janot-Pacheco},
  E. (2001).
\newblock {Interferometric-Doppler imaging of stellar surface structure}.
\newblock {\em \aap\/}, {\bf 377}, 721--734.

\bibitem[{Kervella} {\em et~al.}(2004a){Kervella}, {Nardetto}, {Bersier},
  {Mourard}, and {Coud{\'e} du Foresto}]{kervella04}
{Kervella}, P., {Nardetto}, N., {Bersier}, D., {Mourard}, D., and {Coud{\'e} du
  Foresto}, V. (2004a).
\newblock {Cepheid distances from infrared long-baseline interferometry. I.
  VINCI/VLTI observations of seven Galactic Cepheids}.
\newblock {\em \aap\/}, {\bf 416}, 941--953.

\bibitem[{Kervella} {\em et~al.}(2004b){Kervella}, {Th{\'e}venin}, {Di Folco},
  and {S{\'e}gransan}]{kervella04B}
{Kervella}, P., {Th{\'e}venin}, F., {Di Folco}, E., and {S{\'e}gransan}, D.
  (2004b).
\newblock {The angular sizes of dwarf stars and subgiants. Surface brightness
  relations calibrated by interferometry}.
\newblock {\em \aap\/}, {\bf 426}, 297--307.

\bibitem[{Kervella} {\em et~al.}(2008){Kervella}, {M{\'e}rand}, {Pichon},
  {Th{\'e}venin}, {Heiter}, {Bigot}, {Ten Brummelaar}, {McAlister}, {Ridgway},
  {Turner}, {Sturmann}, {Sturmann}, {Goldfinger}, and {Farrington}]{kervella08}
{Kervella}, P., {M{\'e}rand}, A., {Pichon}, B., {Th{\'e}venin}, F., {Heiter},
  U., {Bigot}, L., {Ten Brummelaar}, T.~A., {McAlister}, H.~A., {Ridgway},
  S.~T., {Turner}, N., {Sturmann}, J., {Sturmann}, L., {Goldfinger}, P.~J., and
  {Farrington}, C. (2008).
\newblock {The radii of the nearby K5V and K7V stars 61 Cygni A \& B.
  CHARA/FLUOR interferometry and CESAM2k modeling}.
\newblock {\em \aap\/}, {\bf 488}, 667--674.

\bibitem[{Kervella} {\em et~al.}(2009){Kervella}, {Domiciano de Souza},
  {Kanaan}, {Meilland}, {Spang}, and {Stee}]{kervella09}
{Kervella}, P., {Domiciano de Souza}, A., {Kanaan}, S., {Meilland}, A.,
  {Spang}, A., and {Stee}, P. (2009).
\newblock {The environment of the fast rotating star Achernar. II. Thermal
  infrared interferometry with VLTI/MIDI}.
\newblock {\em \aap\/}, {\bf 493}, L53--L56.

\bibitem[{Kraus} {\em et~al.}(2007){Kraus}, {Balega}, {Berger}, {Hofmann},
  {Millan-Gabet}, {Monnier}, {Ohnaka}, {Pedretti}, {Preibisch}, {Schertl},
  {Schloerb}, {Traub}, and {Weigelt}]{kraus07}
{Kraus}, S., {Balega}, Y.~Y., {Berger}, J.-P., {Hofmann}, K.-H.,
  {Millan-Gabet}, R., {Monnier}, J.~D., {Ohnaka}, K., {Pedretti}, E.,
  {Preibisch}, T., {Schertl}, D., {Schloerb}, F.~P., {Traub}, W.~A., and
  {Weigelt}, G. (2007).
\newblock {Visual/infrared interferometry of Orion Trapezium stars: preliminary
  dynamical orbit and aperture synthesis imaging of the ${\theta}^{1}$ Orionis
  C system}.
\newblock {\em \aap\/}, {\bf 466}, 649--659.

\bibitem[{Lane} {\em et~al.}(2000){Lane}, {Kuchner}, {Boden}, {Creech-Eakman},
  and {Kulkarni}]{lane2000}
{Lane}, B.~F., {Kuchner}, M.~J., {Boden}, A.~F., {Creech-Eakman}, M., and
  {Kulkarni}, S.~R. (2000).
\newblock {Direct detection of pulsations of the Cepheid star {$\zeta$} Gem and
  an independent calibration of the period-luminosity relation}.
\newblock {\em \nat\/}, {\bf 407}, 485--487.

\bibitem[{MacGregor} {\em et~al.}(2007){MacGregor}, {Jackson}, {Skumanich}, and
  {Metcalfe}]{mac07}
{MacGregor}, K.~B., {Jackson}, S., {Skumanich}, A., and {Metcalfe}, T.~S.
  (2007).
\newblock {On the Structure and Properties of Differentially Rotating,
  Main-Sequence Stars in the 1-2 $M_{\odot}$ Range}.
\newblock {\em \apj\/}, {\bf 663}, 560--572.

\bibitem[{M{\'e}rand} {\em et~al.}(2005){M{\'e}rand}, {Kervella}, {Coud{\'e} Du
  Foresto}, {Ridgway}, {Aufdenberg}, {Ten Brummelaar}, {Berger}, {Sturmann},
  {Sturmann}, {Turner}, and {McAlister}]{merand05}
{M{\'e}rand}, A., {Kervella}, P., {Coud{\'e} Du Foresto}, V., {Ridgway}, S.~T.,
  {Aufdenberg}, J.~P., {Ten Brummelaar}, T.~A., {Berger}, D.~H., {Sturmann},
  J., {Sturmann}, L., {Turner}, N.~H., and {McAlister}, H.~A. (2005).
\newblock {The projection factor of {$\delta$} Cephei. A calibration of the
  Baade-Wesselink method using the CHARA Array}.
\newblock {\em \aap\/}, {\bf 438}, L9--L12.

\bibitem[{M{\'e}rand} {\em et~al.}(2007){M{\'e}rand}, {Aufdenberg}, {Kervella},
  {Foresto}, {ten Brummelaar}, {McAlister}, {Sturmann}, {Sturmann}, and
  {Turner}]{merand07}
{M{\'e}rand}, A., {Aufdenberg}, J.~P., {Kervella}, P., {Foresto}, V.~C.~d.,
  {ten Brummelaar}, T.~A., {McAlister}, H.~A., {Sturmann}, L., {Sturmann}, J.,
  and {Turner}, N.~H. (2007).
\newblock {Extended Envelopes around Galactic Cepheids. III. Y Ophiuchi and
  {$\alpha$} Persei from Near-Infrared Interferometry with CHARA/FLUOR}.
\newblock {\em \apj\/}, {\bf 664}, 1093--1101.

\bibitem[{Minton} and {Malhotra}(2007){Minton} and {Malhotra}]{minton07}
{Minton}, D.~A. and {Malhotra}, R. (2007).
\newblock {Assessing the Massive Young Sun Hypothesis to Solve the Warm Young
  Earth Puzzle}.
\newblock {\em \apj\/}, {\bf 660}, 1700--1706.

\bibitem[{Monnier} {\em et~al.}(2007){Monnier}, {Zhao}, {Pedretti}, {Thureau},
  {Ireland}, {Muirhead}, {Berger}, {Millan-Gabet}, {Van Belle}, {ten
  Brummelaar}, {McAlister}, {Ridgway}, {Turner}, {Sturmann}, {Sturmann}, and
  {Berger}]{monnier07}
{Monnier}, J.~D., {Zhao}, M., {Pedretti}, E., {Thureau}, N., {Ireland}, M.,
  {Muirhead}, P., {Berger}, J.-P., {Millan-Gabet}, R., {Van Belle}, G., {ten
  Brummelaar}, T., {McAlister}, H., {Ridgway}, S., {Turner}, N., {Sturmann},
  L., {Sturmann}, J., and {Berger}, D. (2007).
\newblock {Imaging the Surface of Altair}.
\newblock {\em Science\/}, {\bf 317}, 342--.

\bibitem[{North} {\em et~al.}(2007a){North}, {Tuthill}, {Tango}, and
  {Davis}]{north07}
{North}, J.~R., {Tuthill}, P.~G., {Tango}, W.~J., and {Davis}, J. (2007a).
\newblock {${\gamma}^{2}$ Velorum: orbital solution and fundamental parameter
  determination with SUSI}.
\newblock {\em \mnras\/}, {\bf 377}, 415--424.

\bibitem[{North} {\em et~al.}(2007b){North}, {Davis}, {Bedding}, {Ireland},
  {Jacob}, {O'Byrne}, {Owens}, {Robertson}, {Tango}, and {Tuthill}]{north07B}
{North}, J.~R., {Davis}, J., {Bedding}, T.~R., {Ireland}, M.~J., {Jacob},
  A.~P., {O'Byrne}, J., {Owens}, S.~M., {Robertson}, J.~G., {Tango}, W.~J., and
  {Tuthill}, P.~G. (2007b).
\newblock {The radius and mass of the subgiant star {$\beta$} Hyi from
  interferometry and asteroseismology}.
\newblock {\em \mnras\/}, {\bf 380}, L80--L83.

\bibitem[{Ohishi} {\em et~al.}(2004){Ohishi}, {Nordgren}, and
  {Hutter}]{ohishi04}
{Ohishi}, N., {Nordgren}, T.~E., and {Hutter}, D.~J. (2004).
\newblock {Asymmetric Surface Brightness Distribution of Altair Observed with
  the Navy Prototype Optical Interferometer}.
\newblock {\em \apj\/}, {\bf 612}, 463--471.

\bibitem[{Perrin} {\em et~al.}(2004a){Perrin}, {Ridgway}, {Coud{\'e} du
  Foresto}, {Mennesson}, {Traub}, and {Lacasse}]{perrin04B}
{Perrin}, G., {Ridgway}, S.~T., {Coud{\'e} du Foresto}, V., {Mennesson}, B.,
  {Traub}, W.~A., and {Lacasse}, M.~G. (2004a).
\newblock {Interferometric observations of the supergiant stars {$\alpha$}
  Orionis and {$\alpha$} Herculis with FLUOR at IOTA}.
\newblock {\em \aap\/}, {\bf 418}, 675--685.

\bibitem[{Perrin} {\em et~al.}(2004b){Perrin}, {Ridgway}, {Mennesson},
  {Cotton}, {Woillez}, {Verhoelst}, {Schuller}, {Coud{\'e} du Foresto},
  {Traub}, {Millan-Gabet}, and {Lacasse}]{perrin04}
{Perrin}, G., {Ridgway}, S.~T., {Mennesson}, B., {Cotton}, W.~D., {Woillez},
  J., {Verhoelst}, T., {Schuller}, P., {Coud{\'e} du Foresto}, V., {Traub},
  W.~A., {Millan-Gabet}, R., and {Lacasse}, M.~G. (2004b).
\newblock {Unveiling Mira stars behind the molecules. Confirmation of the
  molecular layer model with narrow band near-infrared interferometry}.
\newblock {\em \aap\/}, {\bf 426}, 279--296.

\bibitem[{Peterson} {\em et~al.}(2006a){Peterson}, {Hummel}, {Pauls},
  {Armstrong}, {Benson}, {Gilbreath}, {Hindsley}, {Hutter}, {Johnston},
  {Mozurkewich}, and {Schmitt}]{peterson06}
{Peterson}, D.~M., {Hummel}, C.~A., {Pauls}, T.~A., {Armstrong}, J.~T.,
  {Benson}, J.~A., {Gilbreath}, G.~C., {Hindsley}, R.~B., {Hutter}, D.~J.,
  {Johnston}, K.~J., {Mozurkewich}, D., and {Schmitt}, H. (2006a).
\newblock {Resolving the Effects of Rotation in Altair with Long-Baseline
  Interferometry}.
\newblock {\em \apj\/}, {\bf 636}, 1087--1097.

\bibitem[{Peterson} {\em et~al.}(2006b){Peterson}, {Hummel}, {Pauls},
  {Armstrong}, {Benson}, {Gilbreath}, {Hindsley}, {Hutter}, {Johnston},
  {Mozurkewich}, and {Schmitt}]{vega06_peterson}
{Peterson}, D.~M., {Hummel}, C.~A., {Pauls}, T.~A., {Armstrong}, J.~T.,
  {Benson}, J.~A., {Gilbreath}, G.~C., {Hindsley}, R.~B., {Hutter}, D.~J.,
  {Johnston}, K.~J., {Mozurkewich}, D., and {Schmitt}, H.~R. (2006b).
\newblock {Vega is a rapidly rotating star}.
\newblock {\em Nature\/}, {\bf 440}, 896--899.

\bibitem[{Quirrenbach} {\em et~al.}(1996){Quirrenbach}, {Mozurkewich},
  {Buscher}, {Hummel}, and {Armstrong}]{quirrenbach96}
{Quirrenbach}, A., {Mozurkewich}, D., {Buscher}, D.~F., {Hummel}, C.~A., and
  {Armstrong}, J.~T. (1996).
\newblock {Angular diameter and limb darkening of Arcturus.}
\newblock {\em \aap\/}, {\bf 312}, 160--166.

\bibitem[{Rieke} {\em et~al.}(2008){Rieke}, {Blaylock}, {Decin}, {Engelbracht},
  {Ogle}, {Avrett}, {Carpenter}, {Cutri}, {Armus}, {Gordon}, {Gray}, {Hinz},
  {Su}, and {Willmer}]{rieke08}
{Rieke}, G.~H., {Blaylock}, M., {Decin}, L., {Engelbracht}, C., {Ogle}, P.,
  {Avrett}, E., {Carpenter}, J., {Cutri}, R.~M., {Armus}, L., {Gordon}, K.,
  {Gray}, R.~O., {Hinz}, J., {Su}, K., and {Willmer}, C.~N.~A. (2008).
\newblock {Absolute Physical Calibration in the Infrared}.
\newblock {\em \aj\/}, {\bf 135}, 2245--2263.

\bibitem[{Schaefer} {\em et~al.}(2008){Schaefer}, {Simon}, {Prato}, and
  {Barman}]{sspb2008}
{Schaefer}, G.~H., {Simon}, M., {Prato}, L., and {Barman}, T. (2008).
\newblock {Preliminary Orbit of the Young Binary Haro 1-14c}.
\newblock {\em \aj\/}, {\bf 135}, 1659--1668.

\bibitem[{Tycner} {\em et~al.}(2006){Tycner}, {Gilbreath}, {Zavala},
  {Armstrong}, {Benson}, {Hajian}, {Hutter}, {Jones}, {Pauls}, and
  {White}]{tycner06}
{Tycner}, C., {Gilbreath}, G.~C., {Zavala}, R.~T., {Armstrong}, J.~T.,
  {Benson}, J.~A., {Hajian}, A.~R., {Hutter}, D.~J., {Jones}, C.~E., {Pauls},
  T.~A., and {White}, N.~M. (2006).
\newblock {Constraining Disk Parameters of Be Stars using Narrowband
  H{$\alpha$} Interferometry with the Navy Prototype Optical Interferometer}.
\newblock {\em \aj\/}, {\bf 131}, 2710--2721.

\bibitem[{Wittkowski} {\em et~al.}(2001){Wittkowski}, {Hummel}, {Johnston},
  {Mozurkewich}, {Hajian}, and {White}]{wittkowski01}
{Wittkowski}, M., {Hummel}, C.~A., {Johnston}, K.~J., {Mozurkewich}, D.,
  {Hajian}, A.~R., and {White}, N.~M. (2001).
\newblock {Direct multi-wavelength limb-darkening measurements of three
  late-type giants with the Navy Prototype Optical Interferometer}.
\newblock {\em \aap\/}, {\bf 377}, 981--993.

\bibitem[{Wittkowski} {\em et~al.}(2004){Wittkowski}, {Aufdenberg}, and
  {Kervella}]{psi_phe04}
{Wittkowski}, M., {Aufdenberg}, J.~P., and {Kervella}, P. (2004).
\newblock {Tests of stellar model atmospheres by optical interferometry.
  VLTI/VINCI limb-darkening measurements of the M4 giant {$\psi$} Phe}.
\newblock {\em \aap\/}, {\bf 413}, 711--723.

\bibitem[{Wittkowski} {\em et~al.}(2008){Wittkowski}, {Boboltz}, {Driebe}, {Le
  Bouquin}, {Millour}, {Ohnaka}, and {Scholz}]{wittkowski08}
{Wittkowski}, M., {Boboltz}, D.~A., {Driebe}, T., {Le Bouquin}, J.-B.,
  {Millour}, F., {Ohnaka}, K., and {Scholz}, M. (2008).
\newblock {J, H, K spectro-interferometry of the Mira variable S Orionis}.
\newblock {\em \aap\/}, {\bf 479}, L21--L24.

\bibitem[{Zuckerman} and {Song}(2004){Zuckerman} and {Song}]{zs2004}
{Zuckerman}, B. and {Song}, I. (2004).
\newblock {Young Stars Near the Sun}.
\newblock {\em \araa\/}, {\bf 42}, 685--721.

\end{thebibliography}
\end{document}